\documentclass[runningheads]{svmult}
\usepackage{makeidx}   % allows index generation
\usepackage{graphicx}  % standard LaTeX graphics tool
\usepackage{subeqnar}  % subnumbers individual equations
\usepackage{multicol}  % used for the two-column index
\usepackage{cropmark} % cropmarks for pages without
\usepackage{physprbb}  % centered layout of diverse elements, etc.
\begin{document}
\title*{Testing A (Stringy) Model of Quantum Gravity} 
\toctitle{Testing A (Stringy) Model of Quantum Gravity} 
% allows explicit linebreak for the table of content
%
%
\titlerunning{Testing Quantum Gravity}
% allows abbreviation of title, if the full title is too long
% to fit in the running head
%
\author{Nick E. Mavromatos}
\authorrunning{Nick E. Mavromatos}
% if there are more than two authors,
% please abbreviate author list for running head
%
%
\institute{Department of Physics, Theoretical Physics,
King's College London, Strand, London WC2R 2LS, United Kingdom, and \\
CERN, Theory Division, CH-1211 Geneva 23, Switzerland}

\maketitle              % typesets the title of the contribution

\begin{abstract}

I discuss a specific model of space-time foam, 
inspired by the modern non-perturbative 
approach to string theory (D-branes). 
The model views our world as a 
three brane, intersecting with D-particles that represent 
stringy quantum gravity effects, which can be real or virtual.
In this picture, matter is represented generically by (closed or open) 
strings on the D3 brane propagating 
in such a background. Scattering of the (matter) 
strings off the D-particles
causes recoil of the latter, which in turn 
results in a distortion of the 
surrounding space-time fluid and 
the formation of (microscopic, i.e. Planckian size) 
horizons around the defects. 
As a mean-field result, 
the dispersion relation of the various particle excitations is 
modified, leading to non-trivial optical properties 
of the space time, for instance 
a non-trivial refractive index for the case of photons or other massless 
probes. Such models make falsifiable predictions,
that may be tested experimentally in the foreseeable 
future. I describe a few such tests, ranging from
observations of light from distant gamma-ray-bursters
and ultra high energy 
cosmic rays, to tests using gravity-wave interferometric devices  
and terrestrial particle physics 
experiments involving, for instance, neutral kaons. 

\end{abstract}

\section{Introduction}

The theory of Quantum Gravity still eludes us, despite considerable 
efforts of theorists for more than fifty years. This may be partly due to the 
fact that the theory by its very nature is associated with the structure of
space and time, and as such it may be completely different 
from theories that describe the rest of the (known) fundamental 
interactions in Nature.
In fact, many of the properties that characterize the known field theories,
such as locality, renormalizability, unitarity etc, may not be features of a 
complete theory of quantum gravity. 
Even symmetries, such as Lorentz invariance, may not be exact
at Planck length scales, $\ell_P \sim 10^{-35}$ m, 
which is the characteristic scale at which 
quantum gravity effects are expected to set in. 

At present, there is no complete mathematical model for  
quantum gravity. However, there 
is  a considerable number of  attempts, which may be classified, roughly, 
into 
three major categories. The first is the {\it canonical approach},
where one tries to formulate the model in a {\it background} independent 
way, i.e. to give space time at Planckian scales 
a polymer-like structure, similar to `spin networks', 
and from these (rather abstract) 
building blocks to construct the interactions~\cite{loop}
and the observable universe.
The second approach is the one in which the quantum gravitational 
interactions
are represented as a `stochastic medium', which gives space time 
non-trivial optical properties (`space-time foam')~\cite{foam}. 
This approach,
which is more phenomenological than the first one, 
is based on the expectation that any consistent field theory of 
quantum gravity should involve microscopic event horizons (foam),
surrounding black holes or other singularities of space time. The latter
are classical solutions of 
Einstein theory (or its extensions), and therefore 
such configurations should also 
represent quantum fluctuations, which should be part of the 
(still unknown) complete 
integration measure of the gravitational path integral.
The third approach, and so far the most developed formally,  
is string theory and its modern non-perturbative 
extension (D-branes)~\cite{strings,Polch}.
The discovery of D branes has revolutionised the study of
black-hole physics. Now one has quasi-realistic string models
of black holes in different dimensions, which one can use to study
profound issues concerning the reconciliation of general relativity
and quantum mechanics. A key breakthrough was the
demonstration that the entropy of a stringy black hole
corresponds to the number of its distinct quantum states~\cite{Sen}.
Thus D branes offer the prospect of accounting exactly for the
flow of information in processes involving particles and
black holes. 

At first instance, the string approach to 
quantum gravity may seem to have overcome 
the loss of unitarity that is believed to characterise 
quantum gravitational interactions in the second (`space-time foam')
approach. 
However it is not immediately apparent that an 
observer in string theory 
will not perceive loss of information in any given
particle/D-brane interaction: the answer depends whether
she/he is able to recover all the information transferred from the
scattering particle to the recoiling black hole.
It is important to address this issue at both the macroscopic
and microscopic levels, where the answers may differ. In the case
of a macroscopic black hole, it is difficult to see how {\it in
practice} all the quantum information may be recovered 
without a complete set of observations of the emitted Hawking
radiation~\cite{Hawking}. However, even if this is possible {\it in
principle},
the problem of the microscopic `end-game' that terminates the
Hawking evaporation process is unsolved, in our view.
In this sense, the string theory approach to quantum gravity 
may lead to effective 
stochastic models. However, such stochastic models usually employ 
{\it non-equilibrium} physics, and as such cannot be described by 
critical string theory. Indeed, as we shall discuss below, 
it is our belief that proper quantum gravitational 
interactions in string theory involve at a certain stage departure
from equilibrium, which, in terms of string-theory nomenclature,
implies~\cite{emn98} 
the inclusion of non-critical (Liouville) strings~\cite{david88,distler89}.
The latter 
quantify the process of information loss in a mathematically
consistent way, as we shall discuss in the next section. 

It may be useful for what follows to recall one of the intuitive ways of
formulating the information loss in the process of Hawking radiation
from a macroscopic black hole, whose stringy analogue we 
discuss in
this talk. Consider the quantum-mechanical creation of a pure-state 
particle pair $\vert A,B \rangle$
close to the (classical) black-hole horizon of such a macroscopic black
hole. One can then envisage that particle $B$ falls inside this
horizon, whilst particle $A$ escapes as Hawking radiation. The quantum
state of the
particle $B$ is apparently unobservable, and hence information is
apparently lost. 

This argument is very naive, and one would like to formulate a
more precise treatment of this process at the microscopic level,
suitable for describing space-time foam~\cite{foam}.
The purpose of this talk is to 
review such a 
specific stringy 
treatment~\cite{kogan96,ellis96,mavro+szabo} of the interaction
between
closed-string particle `probes' and D-brane black holes (defects).
We have developed an approach capable of accommodating the
recoil of a D-brane black hole struck by a closed-string `probe',
including also quantum effects associated with higher-genus
contributions to the string path integral. We have shown
explicitly~\cite{kanti98,ellis96} how the loss of information to the
recoiling
D brane (assuming that it is unobserved) leads to information loss,
for both the scattered particle and also any spectator particle. This
information loss can be related to a change in the background metric
following the scattering event, which can be regarded as creating an
Unruh-like `thermal' state.

In a recent paper~\cite{horizons}, which we shall review in the next section, 
we took this line of argument a step further, by
demonstrating that closed-string particle/D-brane scattering
leads in general to the formation of a microscopic event horizon,
within which string particles may be trapped. The scattering
event causes expansion of this horizon, which is eventually halted
and reversed by Hawking radiation~\cite{Hawking}. Thus we have a
microscopic
stringy realization of this process.
A peculiarity of this approach is that the conformal invariance conditions
select preferentially backgrounds with three spatial dimensions. This
leads to a consistent formulation of the interaction of D3 branes with
recoiling D particles, which are allowed to fluctuate independently only
on the D3-brane hypersurface. Some physical consequences of the model 
and their (possible) experimental tests,
which could be generic to other models of space-time foam, 
will be also reviewed in this talk.

\section{A Stringy Model of Space-Time Foam}

\begin{figure}[b]
\begin{center}
\includegraphics[width=.1\textwidth]{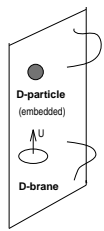}
\end{center}
\caption[]{The world as a D3 brane 
`punctured' by D particles. 
The scattering on the D-particles of string states, either
closed (gravitons) or open (matter fields) that
live on the D3 brane, cause the D-particle
to recoil, leading to stochastic effects in the
propagation of the low-energy states, as well as to 
non-zero `vacuum' energy on the D3 brane.} 
\label{fig3}
\end{figure}

In this section we shall review the basic features of a theoretical model 
of space-time foam, proposed in \cite{horizons}, which is based on 
a modern version of non-critical string theory. 
According to this model, our world is viewed as a 
fluctuating D(irichlet) 3-brane,
embedded in a higher-dimensional (bulk) space time (see 
fig. \ref{fig3}). For formal reasons, we start with a 
Euclideanized $D4$ brane 
and follow the procedure
of identifying the Liouville mode (arising from recoil)
with the (Euclidean) time coordinate $X^0$.
As discussed in \cite{horizons} this procedure
results eventually in a Minkowskian signature.   
The D3 brane
is punctured by D-particles,
that represent defects in the fabric of the D3 space time, 
and thus can be viewed as genuine quantum-gravity singular 
effects (c.f. analogy with microscopic black holes).
The foamy structure of space time 
is caused as a consequence of the distortion of the space-time fluid 
surrounding the 
D-particle, 
which occurs 
due to the recoil of the D-particle 
as a result of the scattering of matter strings off it. 
The recoil of the (massive, hyper-Planckian) defect causes 
the formation of microscopic horizons, which surround the defect, 
trapping part of the scattered matter in the interior. 

In the language of (perturbative) world-sheet ($\sigma$-model) 
string theory, which 
suffices to describe the recoil 
effects to leading order, 
target-space 
quantum fluctuations are incorporated
by appropriate summation over 
world-sheet topologies~\cite{mavro+szabo}. 
Such higher-genus effects 
lead to an oscillating behaviour of the horizon, characterized by 
initial expansion, stasis, and shrinking,
which in physical terms may be interpreted as a result 
of a phenomenon analogous
to Hawking radiation in conventional field theory~\cite{Hawking}.
The above situation involves a single-scattering event. However, 
conceptually one may think of a statistical ensemble of (virtual)
defects, whose scattering with matter strings will create
analogous phenomena involving a statistical distribution 
of dynamical horizons. This is the stringy picture of space time 
foam, whose consequences we shall explore in this talk.

\subsection{Formulation of D-Brane Recoil}

We now proceed to review briefly the mathematical formalism 
underlying the above model.
As discussed in references~\cite{kogan96,ellis96,mavro+szabo}, the
recoil of a D-brane string soliton after
interaction with a closed-string state is
characterized
by a $\sigma$ model on the string world sheet $\Sigma$, that is deformed
by a pair of logarithmic operators~\cite{lcft}:
\begin{equation}
C^I_\epsilon = \epsilon \Theta_\epsilon (X^I),\qquad
D^I_\epsilon = X^I \Theta_\epsilon (X^I), \qquad I \in \{0,\dots, 3\}
\label{logpair}
\end{equation}
defined on the boundary $\partial \Sigma$ of the string world
sheet. Here $X^I, I\in \{0, \dots, p\}$ obey Neumann boundary
conditions on $\Sigma$, and denote the D-brane
coordinates, whilst $\epsilon\rightarrow0^+$ is
a regulating parameter and $\Theta_{\epsilon}(X^{I})$ is a
regularized Heaviside step function.
The remaining $y^i, i\in \{p+1, \dots, 9\}$ in (\ref{logpair}) denote the
transverse bulk directions (c.f. fig. \ref{fig3}).
For reasons of convergence of the world-sheet path integrals
we take the space-time $\{ X^I, y^i \}$ to have {\it Euclidean} 
signature.

In the case of D particles~\cite{kogan96,ellis96,mavro+szabo},  
the index $I$ takes the
value $0$ only, 
in which case the operators (\ref{logpair}) act as
deformations of the conformal field theory on the world sheet. The
operator $u_i \int _{\partial \Sigma} \partial_n X^i D_\epsilon$
describes the movement of the D brane induced by the scattering,
where $u_i$ is its recoil velocity, and $y_i \int _{\partial
\Sigma} \partial_n X^i C_\epsilon $ describes quantum fluctuations
in the initial position $y_i$ of the D particle. It has been
shown rigorously~\cite{mavro+szabo} that the logarithmic conformal
algebra ensures energy--momentum  conservation during the recoil
process: $u_i = ( k^1_i + k^2_i)/M_D$, 
where $k^1 (k^2)$ is
the momentum of the propagating closed string state before (after)
the recoil, and $M_D=1/(\ell _s g_s)$ is the mass of the D brane,
where $g_s$ is the string coupling, which is assumed
here to be weak enough to ensure that the D brane is very massive,
and $\ell _s$ is the string
length.

In order to realize the logarithmic
algebra between the operators $C$ and $D$ (\ref{logpair}), 
one uses
as a regulating parameter~\cite{kogan96}
\begin{equation}
\epsilon^{-2} \sim \ln [L/a] \equiv \Lambda,
\label{defeps}
\end{equation}
where $L$ ($a$) is an infrared (ultraviolet) world--sheet cutoff.
The recoil operators (\ref{logpair}) are
relevant, in the sense of the renormalization group for the
world--sheet field theory, having small conformal dimensions
$\Delta _\epsilon = -\epsilon^2/2$. Thus the $\sigma$-model
perturbed by these operators is not conformal for $\epsilon \ne
0$, and the theory requires Liouville
dressing~\cite{david88,distler89,ellis96}. The consistency of this
approach is supported by the above-mentioned proof of momentum
conservation during the scattering process~\cite{mavro+szabo}.

As discussed in~\cite{ellis96,kanti98}, the recoil deformations
 create a local distortion of the space-time
surrounding the recoiling D brane, which may also be
determined using the method of Liouville dressing.
In~\cite{ellis96,kanti98} we concentrated on describing the
resulting space-time in the case when a D-particle defect embedded in
a $D$-dimensional space-time recoils after the scattering of a
closed string. To leading order in
the recoil velocity $u_i$ of the D particle, the resulting
space-time was found, for times $t \gg 0$ long after the
scattering event at $t=0$, to be equivalent to a Rindler wedge,
with apparent `acceleration' $\epsilon u_i$~\cite{kanti98}, where
$\epsilon$ is defined above (\ref{defeps}).
For times $t < 0$, the space-time is flat Minkowski. There is
hence a discontinuity at $t =0$, which leads to particle
production and decoherence for a low-energy spectator field theory
observer who performs local scattering experiments
long after the scattering, and far away from the
location of the collision of the closed string with the
D particle~\cite{kanti98}.

This situation is easily generalized to D$p$ branes~\cite{emw99}.
The folding/recoil deformations of the D$p$ brane
are relevant deformations, with anomalous dimension
$-\epsilon^2/2 $, which disturbs the conformal invariance of the
world-sheet $\sigma$ model, and restoration of conformal invariance
again requires
Liouville dressing~\cite{david88,distler89,ellis96}, as discussed above.
To determine the effect of
such dressing on the space-time geometry, it is essential  to
write~\cite{ellis96} the boundary recoil deformations as  bulk
world-sheet deformations
\begin{equation}
\int _{\partial \Sigma} {\overline g}_{Iz} x \Theta_\epsilon (x)
\partial_n z =
\int _\Sigma \partial_\alpha \left({\overline g}_{Iz} x \Theta_\epsilon
(x)
\partial ^\alpha z \right)
\label{a1}
\end{equation}
where the ${\overline g}_{Iz}$ denote renormalized
folding/recoil couplings~\cite{mavro+szabo}. Such
couplings are marginal on a flat world sheet, and
the operators (\ref{a1}) are marginal also on a curved
world sheet, provided~\cite{distler89} one dresses the (bulk)
integrand by multiplying it by a factor $e^{\alpha_{Ii}\phi}$,
where $\phi$ is the Liouville field and $\alpha_{Ii}$ is the
gravitational conformal dimension. This is related to the
flat-world-sheet anomalous dimension $-\epsilon^2/2$ of the recoil
operator, viewed as a bulk world-sheet deformation
by~\cite{distler89}: $\alpha_{Ii}=-\frac{Q_b}{2} +
\sqrt{\frac {Q_b^2}{4} + \frac {\epsilon^2}{2} }$,
where $Q_b$ is the central-charge deficit of the bulk world-sheet
theory. In the recoil problem at hand, as discussed
in~\cite{kanti98},
$Q_b^2 \sim \epsilon^4/g_s^2  > 0$, 
for weak folding deformations $g_{Ii}$, and hence one is
confronted with a {\it supercritical} Liouville theory. This
implies a {\it Minkowskian-signature} Liouville-field kinetic term
in the respective $\sigma$ model~\cite{aben89}, which prompts one
to interpret the Liouville field as a time-like target
field. 

There are two approaches which one can follow. 
In the first of them~\cite{leonta}, 
this time is considered as a {\it second} time
coordinate~\cite{emn98}, which is independent of the
(Euclideanized) $X^0$. The presence of this second `time'
does not affect physical observables, which are defined
for appropriate slices with fixed Liouville coordinate, e.g., $\phi
\rightarrow \infty$ or equivalently
$\epsilon \rightarrow 0$.
From the expression for $Q$ we conclude
that $\alpha_{Ii} \sim \epsilon $, to leading order
in perturbation theory in $\epsilon$, to which we restrict
ourselves here. 
In the second approach~\cite{emn98}, which we shall 
mainly follow here, the (Minkowskian) Liouville 
field $\phi$ is identified with the (initially Euclidean) coordinate
$X^0$, and hence one is no longer considering constant Liouville
field slices. In this approach, however, one still identifies 
$\epsilon^{-2}$ with the target time, which in turn implies that 
the perturbative world-sheet approach is valid,
provided one works with sufficiently large times $t$,
i.e. small $\epsilon^2$. 

The $X^I$-dependent field operators
$\Theta_\epsilon (X^I)$ scale with $\epsilon$ as~\cite{ellis96}:
$\Theta_\epsilon(X^I) \sim e^{-\epsilon X^I}
\Theta(X^I)$, where $\Theta(X^I)$ is a Heaviside step function
without any field content, evaluated in the limit $\epsilon \rightarrow 0^+$.
The bulk deformations, therefore, yield the following
$\sigma$-model terms:
\begin{equation}
\frac{1}{4\pi \ell_s^2}~\int _\Sigma 
\sum_{I=0}^{3} \left( \epsilon^2 {\overline g}^C_{Ii} + \epsilon 
{\overline g}_{Ii} X^I\right)
e^{\epsilon(\phi_{(0)} - X^I_{(0)})}\Theta(X^I_{(0)})
\partial_\alpha \phi 
\partial^\alpha y_i~
\label{bulksigma}
\end{equation}
where the subscripts $(0)$ denote world-sheet zero modes, and 
${\overline g}^C_{0i}=y_i$.

Upon the interpretation of the Liouville zero mode $\phi_{(0)}$ as
a (second)
time-like coordinate, the deformations (\ref{bulksigma}) yield
metric
deformations of the generalized space-time
with two times. The metric components
for fixed Liouville-time slices can be
interpreted~\cite{ellis96}
as expressing the distortion of the space-time
surrounding the recoiling D-brane soliton.
For clarity,
we now drop the subscripts $(0)$ for the rest of this paper,
and we work in a region of space-time
such that $\epsilon (\phi - X^I)$ is finite
in the limit $\epsilon \rightarrow 0^+$.
The resulting space-time distortion is therefore
described by the metric elements
\begin{eqnarray}
&~& G_{\phi\phi} = -1, \qquad G_{ij} =\delta_{ij}, \qquad
G_{IJ}=\delta_{IJ}, \qquad G_{iI}=0,   \nonumber \\
&~& G_{\phi i} = \left(\epsilon^2 {\overline g}^C_{Ii} +
 \epsilon {\overline g}_{Ii}X^I \right)\Theta (X^I)~,
\qquad i=4, \dots 9,~~I=0, \dots 3
\label{gemetric}
\end{eqnarray}
where the index $\phi $ denotes Liouville `time', not to be confused
with the Euclideanized time which is one of the $X^I$.
To leading order in $\epsilon {\overline g}_{Ii}$,
we may ignore the $\epsilon^2 {\overline g}^C_{Ii}$ term.
The presence of $\Theta(X^I)$ functions and
the fact that we are working in the region $y_i >0$
indicate that
the induced space-time is piecewise continuous (the
important implications for non-thermal particle production
and decoherence for a spectator low-energy field theory
in such space-times were discussed in~\cite{kanti98,ellis96}, where
the D-particle recoil case was considered).

We next study in more detail some physical aspects of
the metric (\ref{gemetric}),
restricting ourselves, for simplicity, to the case
of a single Dirichlet dimension $z$ that
plays the r\^ole of a bulk dimension
in a set up where there are 
Neumann coordinates $X^I$, $I=0,\dots3$
parametrizing a D4 (Euclidean) brane, interpreted as
our four-dimensional space-time.
Upon performing the time transformation
$\phi \rightarrow \phi - \frac{1}{2}\epsilon {\overline g}_{Iz} X^I z $, the
line element (\ref{gemetric}) becomes:
\begin{eqnarray}
&~&ds^2 =-d\phi^2 + \left(\delta_{IJ}
-\frac{1}{4}\epsilon^2{\overline g}_{Iz}{\overline g}_{Jz}~z^2\right)~dX^I dX^J
+  
\nonumber \\
&~& \left(1 + \frac{1}{4}\epsilon ^2
{\overline g}_{Iz}{\overline g}_{Jz}~X^I~X^J\right)~dz^2 -
\epsilon {\overline g}_{Iz}~z~dX^I~d\phi~, \nonumber \\
\label{bendinglineel}
\end{eqnarray}
where $\phi$ is the Liouville field which, we remind the reader,
has Minkowskian signature in the case of supercritical
strings that we are dealing with here.
One may now make a general coordinate transformation on the
brane $X^I$ that diagonalizes the pertinent induced-metric
elements in (\ref{bendinglineel}) (note that general
coordinate invariance is assumed to be a good symmetry on the
brane, away from the `boundary' $X^I=0$). 
The so-diagonalized metric becomes~\cite{leonta,horizons} 
\begin{eqnarray}
&~&ds^2 =-d\phi^2 + \left(1
-\alpha^2 ~z^2\right)~(dX^I)^2
+ \left(1 + \alpha^2 ~(X^I)^2\right)~dz^2 - \epsilon {\overline g}_{Iz}~z~dX^I~d\phi~,
\nonumber \\
&~& \alpha=\frac{1}{2}\epsilon {\overline g}_{Iz} \sim g_s |\Delta P_z|/M_s
\label{bendinglineel3}
\end{eqnarray}
where the last expression is a reminder that
one can express the parameter
$\alpha$ (in the limit $\epsilon \rightarrow 0^+$)
in terms of the (recoil) momentum transfer $\Delta P_z$ along the bulk
direction.

A last comment, which is important for our purposes here, 
concerns the case in which
the metric (\ref{bendinglineel3}) is {\it exact}, i.e., it holds
to all orders in ${\overline g}_{Iz}z$.
This is the case where
there is no world-sheet
tree-level momentum transfer. This naively corresponds to the case
of static intersecting branes. However, the whole philosophy of
recoil~\cite{kogan96,mavro+szabo} implies that, even in that case,
there are quantum fluctuations induced by the sum over genera of the
world sheet. The latter implies the existence of a statistical
distribution of logarithmic deformation couplings of Gaussian type
about a mean-field value ${\overline g}^{C}_{Iz}=0$. Physically,
the couplings
${\overline g}_{Iz}$ represent recoil velocities of the intersecting
branes,
hence these Gaussian fluctuations
represent the effects of quantum fluctuations about the
zero recoil-velocity case, which may be considered as quantum
corrections to the static intersecting-brane case.
We therefore consider a statistical average
$<< \cdots >>$ of the line element (\ref{bendinglineel})
where $<< \cdots >>=\int _{-\infty}^{+\infty}d{\overline g}_{Iz}
\left(\sqrt{\pi}\Gamma \right)^{-1} 
 e^{-{\overline g}_{Iz}^2/\Gamma^2} (\cdots)$, 
and the width $\Gamma$ 
is
found~\cite{mavro+szabo} after summation over world-sheet
genera to be
proportional to the string coupling $g_s$. 
In fact, it can be shown~\cite{mavro+szabo} that $\Gamma$ scales 
as $\epsilon {\overline \Gamma}$, where ${\overline \Gamma}$ 
is independent of $\epsilon$.

Assuming that $g_{Iz}={\cal O}(|u_i|)$ where $u_i=g_s \Delta P_i/M_s$
is the recoil velocity~\cite{kogan96,mavro+szabo}, we see that
the average line element
$ds^2$ becomes:
\begin{eqnarray}
&~&<<ds^2>> =-d\phi^2 + \left(1
-\alpha^2 ~z^2\right)~(dX^I)^2
+ \left(1 + \alpha^2 ~(X^I)^2\right)~dz^2,
\nonumber \\
&~& \alpha=\frac{1}{2\sqrt{2}}\epsilon^2 {\overline \Gamma}
\label{bendinglineel3ab}
\end{eqnarray}
Thus the average over quantum fluctuations leads to a
metric of the form (\ref{bendinglineel3}), but with a parameter
$\alpha$ determined by the width (uncertainty)
of the pertinent quantum fluctuations~\cite{mavro+szabo}.

An important feature of the line elements (\ref{bendinglineel3}) 
and (\ref{bendinglineel3ab}) 
is
the existence of a {\it horizon} at $z=1/\alpha$ for {\it Euclidean}
Neumann coordinates $X^I$. Since the Liouville field
$\phi$ has decoupled after the averaging procedure, 
one may consider slices of this field, defined by $\phi$ = const, on
which  the physics of the observable world can be studied~\cite{leonta}. 
From a
world-sheet renormalization-group view point this slicing
procedure corresponds to selecting a specific point in the
non-critical-string theory space. Usually, the infrared fixed
point
$\phi \rightarrow \infty$ is selected. In that case
one considers (\ref{defeps}) a slice for which $\epsilon^2 \rightarrow 0$.
But any other choice could do, so $\alpha$ may be considered
a small but arbitrary parameter of our effective theory.
The presence of a horizon raises the issue of how one
could analytically continue so as to pass to the space beyond the horizon.
The simplest way, compatible, as we discussed in
\cite{leonta}, with the low-energy
Einstein's equations, is to take the absolute value of $1-\alpha^2 z^2$
in the metric element (\ref{bendinglineel3}) and/or (\ref{bendinglineel3ab}).

We now pass onto the second approach~\cite{emn98},
in which one identifies the Liouville mode $\phi$ with 
the time coordinate $X^0$ on the initial $Dp$ brane. 
In this case, as we shall see, the situation becomes much more 
interesting, at least  
in certain regions of the bulk space time,
where one can calculate reliably in a world-sheet
perturbative approach. Indeed,
far away from the horizon at $|z|=1/\alpha$, 
i.e., for $\alpha ^2 z^2 << 1$, 
the line element corresponding to the space-time 
(\ref{bendinglineel3ab}) after the identification 
$\phi =X^0$ becomes:
\begin{equation} 
ds^2 \simeq 
-\alpha ^2 z^2 \left(dX^0\right)^2 + dz^2 + \sum_{i=1}^{3}\left(dX^i\right)^2 
\label{conical}
\end{equation} 
implying that $X^0$ plays now the r\^ole of a {\it Minkowskian}-signature
temporal variable, despite its original Euclidean nature. 
This is a result of the identification $\phi=X^0$, and the fact that 
$\phi$ appeared with Minkowskian signature due to the supercriticality 
($Q^2 >0)$ of the Liouville string under consideration. 

Notice that the space time (\ref{conical}) is flat, and hence it satisfies
Einstein's equations, formally. However,
the space time (\ref{conical}) has a {\it conical} singularity  
when one compactifies the time variable $X^0$ on a circle of finite radius
corresponding to 
an inverse `temperature' $\beta$. Formally, 
this requires a Wick rotation 
$X^0 \rightarrow iX^0$ and then compactification, $iX^0=\beta e^{i\theta}$,
$\theta \in \left(0 , 2\pi \right]$.
The space-time then becomes a {\it conical} space-time of Rindler type
\begin{equation} 
  ds_{conical}^2 = \frac{1}{4\pi^2}\alpha^2 \beta^2 z^2 \left( d\theta \right)^2 + dz^2 
+ \sum_{i=1}^{3}\left(dX^i\right)^2 
\label{conical2}
\end{equation} 
with deficit angle $\delta \equiv 2\pi - \alpha \beta$. 
We recall that there is a 
`thermalization theorem' for this space-time~\cite{unruh}, in the 
sense that the deficit 
disappears and the space-time becomes regular, when the temperature 
is fixed to be 
\begin{equation}
     T = \alpha /2\pi  
\label{unruh}
\end{equation} 
The result (\ref{unruh}) may be understood physically 
by the fact that $\alpha$ is essentially related to recoil. As discussed
in \cite{kanti98}, the problem 
of considering a suddenly fluctuating (or recoiling) brane at $X^0=0$,
as in our case above, 
becomes equivalent to that of an observer in a (non-uniformly)
accelerated frame. At times long after the collision the acceleration 
becomes uniform and equals $\alpha$. This implies the appearance of a
non-trivial  
vacuum~\cite{unruh}, characterized by thermal properties of the form 
(\ref{unruh}). At such a temperature the vacuum becomes 
just the Minkowski vacuum, whilst the Unruh vacuum~\cite{unruh} 
corresponds to $\beta \rightarrow \infty$. 
Here we have derived this result in a
different way than in \cite{kanti98}, but the essential physics is 
the same.

\subsection{D-Particle Recoil, Vacuum Energy and the Dimensionality of the
Brane World}

As we have seen above, the recoil of a D-particle in the 
situation of fig. \ref{fig3} induced a non-trivial distortion 
of the D3 hypersurface.
The distortion is such as to induce
non-trivial contributions to the 
vacuum energy on the D3 brane, as discussed in 
detail in~\cite{emncosmol,ellis98,horizons}. 

To see this, we recall that the four-dimensional space-time,
in which the defect is embedded, is to be viewed as a bulk space-time
from the point of view of the world-sheet approach to the recoil of 
the D particle. Following the same approach as that leading to 
(\ref{conical}), involving the identification of the 
Liouville field with the target time, $t$, 
one observes again that there exists an (expanding) horizon, located at 
\begin{equation} 
r^2_h \equiv x_1^2 + x_2^2 + x_3^2 =t^2 /b'^2 
\label{encon}
\end{equation} 
where $\{ x_i \}, i=1,\dots 3$
constitute the bulk dimensions, obeying Dirichlet boundary conditions
on the world sheet, 
and $b'$ is related to the momentum uncertainty of the fluctuating 
D particle. The variance $b'$ was computed~\cite{mavro+szabo}  
using a world-sheet formalism resummed over pinched 
annuli, which has been argued to be the leading-order 
effect for weak string coupling $g_s$:
\begin{equation}
(b')^2 = 4\frac{g_s^2}{\ell_s^2}\left(1 - \frac{285}{18}g_s^2 
\frac{E_{kin}}{M_Dc^2}\right)
+ {\cal O}(g_s^6) 
\label{varianceb'}
\end{equation}
where $E_{kin}$ is the kinetic energy scale of the fluctuating (heavy)
D particle, $M_D=g_s/\ell_s$ is the D--particle mass scale, 
and $\ell_s$ is the string length. Note
the dependence of the variance $b'$ 
on the string coupling $g_s$, which arises because
quantum corrections come from the summation over world-sheet
topologies~\cite{emn98,mavro+szabo}, 
and $g_s$ is a string-loop 
counting parameter.

For the region of space-time
{\it inside the horizon} one obtains the following  metric 
on the D3 brane, as a result of recoil of the D particle
embedded in it:
\begin{equation}
ds^{2{(4)}} \simeq 
\frac{b'^2 r^2}{t^2} \left(dt\right)^2 -  
\sum_{i=1}^{3}\left(dx^i\right)^2~,\qquad r^2 = \sum_{i=1}^3 x_i^2  
< t^2/b'^2
\label{fourdime2}
\end{equation} 
Note that the scalar curvature corresponding to the metric 
(\ref{fourdime2}) 
has the form $R=-4/r^2$, and as such has a singularity 
at the initial 
location 
$r=0$ of the D-particle defect, as expected. 
This metric is a solution of 
Einstein's  equations in a four-dimensional space-time $\{x_i, t \}$,
with  
a non-trivial ``vacuum'' energy $\Lambda$, 
provided there exists 
a four-dimensional dilaton field of the form:
\begin{equation}
  \varphi = {\rm ln}r + b'{\rm ln}t  
\label{lineardila}
\end{equation} 
which has non-trivial potential 
$V(\varphi)$:  
\begin{equation}
V(\varphi_c)=\frac{1}{r^2}~,\qquad \Lambda =\frac{1}{r^2} 
\label{vlambda}
\end{equation} 
Above we have ignored the fluctuations 
of the D3 brane in the bulk directions. When these are taken into 
account there may be additional contributions to the 
vacuum and excitation energies on the D3 brane, 
which in fact are time-dependent, relaxing to zero asymptotically,
as discussed in 
\cite{emn98,adrian+mavro}.  

From (\ref{vlambda}),(\ref{lineardila}) we observe that, in general,  
there is an explicit $r$ or $t$ dependence in the dilaton potential $V$ 
and the vacuum energy $\Lambda$, which cannot be all absorbed in 
the field $\varphi$. 
This implies {\it violation of 
Lorentz invariance } in the bulk, as a result of the recoil of the 
D3 brane. The result is in agreement with the thermalization
theorem (\ref{unruh}), discussed in subsection 2.1. 

It is interesting to remark~\cite{horizons} 
that 
the metric 
equations 
are satisfied for
the simple case of a free scalar (dilaton) field $\varphi$ 
of the form (\ref{lineardila}), {\it only for} 
$d=3$ spatial Neumann coordinates,  
independent of the value of $b'$. 
It seems therefore 
that the restoration of conformal 
invariance in the case of recoiling D particles embedded
in a $Dp$ brane, or equivalently the satisfaction of the 
corresponding equations of motion in the Liouville-dressed problem,  
constrains the number of longitudinal 
dimensions on the $Dp$ brane to three.
In other words, {\it only a D3 brane can intersect 
with recoiling (fluctuating) D particles}
in a way consistent with the restoration of conformal invariance
in the manner explored here.
 
\subsection{Energy Conditions and Horizons in Recoil-Induced Space-Times}

It is interesting to look at the energy conditions 
of such space times, which would determine whether
ordinary matter can exist within the horizon region displayed above.
There are various forms of energy 
conditions~\cite{energycond}, which may be expressed as follows: 
\begin{eqnarray} 
{\rm Strong}~~~ &:&~~~ \left(T_{\mu\nu}-\frac{1}{D-2}
g_{\mu\nu}T_\alpha^\alpha\right)\xi^\mu\xi^\nu \ge 0, \nonumber \\
{\rm Dominant}~~~&:& ~~~ T_{\mu\nu}\xi^\mu\eta^\nu \ge 0, \nonumber \\
{\rm Weak}~~~&:& ~~~T_{\mu\nu}\xi^\mu\xi^\nu \ge 0, \nonumber \\
{\rm Weaker}~~~&:& ~~~T_{\mu\nu}\zeta^\mu\zeta^\nu \ge 0.
\label{energyconds}
\end{eqnarray}
where $g_{\mu\nu}$ is the metric and $T_{\mu\nu}$ the 
stress-energy tensor in a $D$-dimensional space-time,
including vacuum-energy contributions,   
$\xi^\mu$ and $\eta^\mu$ are arbitrary future-directed time-like
or null vectors, and $\zeta^\mu$ is an arbitrary null vector. 
The conditions (\ref{energyconds}) have been listed in decreasing
strength, in the sense that
each condition is implied by all its preceding
ones. 

It can be easily seen from Einstein's equations for the metric
(\ref{fourdime2}) 
that inside the horizon $b'^2r^2 \le t^2$ the 
conditions are satisfied, which implies that stable matter can 
exist  {\it inside} such regions of the recoil space-time. 
On the other hand, {\it outside the horizon} 
the recoil-induced metric assumes the form: 
\begin{equation}
ds^{2{(4)}} \simeq 
\left(2 - \frac{b'^2 r^2}{t^2}\right) \left(dt\right)^2 -  
\sum_{i=1}^{3}\left(dx^i\right)^2~,\qquad r^2 > t^2/b'^2
\label{fourdime2b}
\end{equation} 
The induced scalar curvature is easily found to be:
$$R=-4b'^2\left(-3t^2 + b'^2r^2 \right)/\left(-2t^2 + b'^2r^2\right)^2~.$$ 
Notice that there is a {\it curvature} singularity 
at $2t^2 =b'^2r^2$, which is precisely  
the point where there is a signature change in the metric 
(\ref{fourdime2b}).

Notice also that, 
in order to ensure 
a Minkowskian signature in the space-time (\ref{fourdime2b}),
one should impose the restriction  
$2 > \frac{b'^2r^2}{t^2} > 
1.$ Outside this region, the metric becomes {\it Euclidean},
which matches our (formal) construction of having initially a (static)
Euclidean D4 brane embedded in a higher-dimensional (bulk) space time.  
Notice that in such a region one can formally pass onto a Minkowskian 
four-dimensional space time upon a Wick rotation of the (Euclidean) 
time coordinate $X^0$. In this (Wick-rotated) framework, then, the 
space time inside the bubbles
retains its Minkowskian signature due to the specific 
form of the metric (\ref{fourdime2}). 

\begin{figure}[b]
\begin{center}
\includegraphics[width=.4\textwidth]{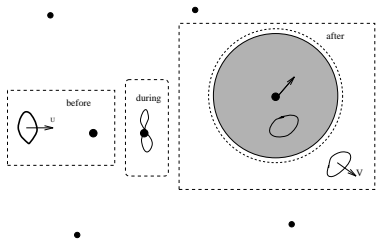}
\end{center}
\caption[]{A schematic representation of scattering in D-foam
background. The dashed boxes represent events just before, during
and after the scattering of a closed-string probe on one particular
D brane defect. The scattering results in the formation of
an shaded bubble, expanding as indicated by the dotted line, inside
which matter can be trapped and
there is an energy-dependent refractive index.}
\label{figbubble}
\end{figure}

The above metric (\ref{fourdime2b}) does not satisfy simple 
Einstein's equations, but this was to be expected,
since the formation of such space-times is not necessarily a 
classical phenomenon. 
In ref. \cite{horizons}  we linked this fact with the failure 
of the energy conditions in this exterior geometry. 
These considerations suggest that matter can  be trapped  
{\it inside} such horizon regions around a fluctuating D-particle defect. 
This sort of trapping is interesting
for our space-time-foam picture, as it implies that such
{\it microscopic D-brane  
horizons } act in a similar way as the intuitive description of a
{\it macroscopic black-hole horizon } discussed in the Introduction, 
as illustrated in Fig.~\ref{figbubble}.

To reinforce the interpretation that
matter is trapped in the interior of a region
described by the metric (\ref{fourdime2}),
we now show that a matter probe inside the horizon `bubble'
experiences an energy-dependent velocity of light.
First rewrite the metric in a Friedmann-Robertson-Walker (FRW) form: 
\begin{equation} 
    ds^2 =e^{2{\rm ln}r} \left(b'^2 dt_{FRW}^2 - \frac{1}{r^2}
\sum_{i=1}^{3} (dx_i)^2 \right)
 \label{scalefactors}
\end{equation} 
where we 
were careful when performing coordinate redefinitions 
{\it not} to absorb in them the factor $b'$, which,  
depends (\ref{varianceb'}) on the energy scale of the matter probe.
We are interested in  matter at various energies
propagating {\it simultaneously} in such a space-time,
and performing a coordinate transformation  
could not absorb an energy-dependent factor such as $b'$.
When we consider the encounter of a matter probe,
such as a photon, with a fluctuating D-particle defect, 
the kinetic-energy scale $E_{kin}$ may be identified with the
energy scale $E$ of the matter probe. We recall
that energy conservation has been proven rigorously
in the world-sheet approach to D--brane recoil~\cite{mavro+szabo},
and survives the resummation over higher genera.
   
We observe from (\ref{scalefactors}) that the overall scale
factor may be absorbed 
into a redefinition of the spatial part of the dilaton 
(\ref{lineardila}), implying that
stable matter experiences an energy-dependent `light velocity' 
\begin{equation}
  c_{int}(E) = b'c = 2cg_s\left(1 - \frac{285g^2_sE}{18M_Dc^2}\right)^{1/2} 
\label{refrindex}
\end{equation}
in the space-time (\ref{fourdime2}),
where $M_D=M_s/g_s$ is the D-particle mass scale. 
The energy-independent factor $2g_s$ may in fact be absorbed
into the normalization of the FRW time coordinate $t_{FRW}$,
thereby making a smooth connection with the velocity of light {\it in
vacuo} in the limiting case of $E/M_Dc^2 \rightarrow 0$.  
{\it It is important to note that, because of the specific form
(\ref{varianceb'}) 
of the variance $b'$, the resulting effective velocity (\ref{refrindex})  
in the interior of the bubble is subluminal~\cite{ellis99}}. 
On the other hand, we see from (\ref{fourdime2b}) that
matter propagates at the normal {\it in vacuo} light velocity 
$c$ in the exterior part of the geometry.

If one considers pulses containing many photons of different
energies~\cite{sarkar,efmmn}, then 
the various photons will experience, as a 
result of the dynamical formation of horizons, changes in 
their mean effective velocities corresponding on average to a
{\it refractive index} $\Delta c(E)$, where
the effective light velocity:
\begin{equation}
c(E)=c\left(1 - \xi \frac{g_sE}{M_sc^2} \right).
\label{refind}\end{equation} 
Here $\xi$ is a quantity that depends on the actual details of the 
scenario for quantum space-time foam, in particular on the density of the
D-brane defects in space. 
In a dilute-gas approximation, $\xi$ might plausibly be assumed to be of
order one, as can be seen as follows.
Consider a path $L$ of a photon, which encounters ${\cal N}$ 
fluctuating D-particle defects. Each defect creates a bubble
which is expected to be close to the Planckian size $\ell_s$, for any 
reasonable model of space-time foam.
Inside each bubble, the photon propagates with velocity (\ref{refrindex}),
whereas outside it propagates with the velocity of light in vacuo $c$.
The total time of flight for this probe will therefore be given by: 
\begin{equation} 
t_{total} = \frac{L-{\cal N}\ell_s}{c} + {\cal N} \frac{\ell_s}{c}
\left(1 - g_s^2\frac{285}{18}\frac{E}{M_Dc^2}\right)^{-1/2}
\label{flighttime}
\end{equation}
In a `dilute gas approximation' for the description of space time 
foam, it is natural to assume that a photon encounters, on average,  
${\cal O}(1)$ D particle
defect in each Planckian length $\ell_s$, so that
${\cal N} \sim \xi L/\ell_s$, where $\xi \le 1$.
From (\ref{flighttime}), then, one obtains a delay in the arrival time 
of a photon of order
\begin{equation} 
\Delta t \sim \xi g_s^2 \frac{285}{36}\frac{L~E}{M_Dc^3} + \dots,
\label{finaldelay}
\end{equation}
corresponding to the effective velocity (\ref{refind}). In
conventional 
string theory, $g_s^2/2\pi \sim 1/20$, and the overall numerical factor
in (\ref{finaldelay}) is of order $4.4~\xi$. However, $g_s$ 
should rather be considered an arbitrary parameter of the model, which may
then 
be constrained by phenomenological observations~\cite{efmmn} through 
limits on (\ref{finaldelay}).

\subsection{Breathing Horizons in Liouville String Theory and 
the emergence of Space-Time Foam}

The tendency of the horizon (\ref{encon}) 
to expand is a classical 
feature. Upon quantization, which corresponds
in our picture to a proper resummation
over world-sheet topologies, one expects a phenomenon similar to Hawking 
radiation. Such a phenomenon would decelerate and stop the expansion,
leading eventually to the shrinking of the horizon. 
This would be a dynamical picture 
of space-time foam, which unfortunately at present is not fully available,
given that at microscopic distances the world-sheet perturbative 
analysis breaks down. However, we believe 
that this picture is quite plausible, and
we can support these considerations formally
by recalling that time $t$ is the 
Liouville field in our formalism. 

\begin{figure}[b]
\begin{center}
\includegraphics[width=.3\textwidth]{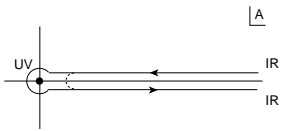}
\end{center}
\caption[]{Contour
of integration appearing in the analytically-continued
(regularized) version of world-sheet Liouville string correlators.
The quantity $A$ denotes the (complex) world-sheet area. 
This is known in the literature as the Saalschutz contour~\cite{ctp}, and
has been used in
conventional quantum field theory to relate dimensional
regularization to the Bogoliubov-Parasiuk-Hepp-Zimmermann
renormalization method. Upon the identification of the 
Liouville field with target time, this curve
resembles closed-time paths in non-equilibrium field theories.}
\label{bounce}
\end{figure}

The dynamics of the
Liouville field exhibits a `bounce' behaviour, when considered
from a world-sheet view point~\cite{kogan,emn98}, as illustrated 
in Fig.~\ref{bounce}. 
This is a general feature of non-critical strings,
whenever the Liouville field is viewed as a local 
renormalization-group scale
of the world sheet. 
The flow of the Liouville scale is 
in both directions between fixed points of the 
world-sheet renormalization group: $ {\rm Infrared} ~ {\rm fixed} ~ 
{\rm point}  \rightarrow {\rm  Ultraviolet} ~ {\rm fixed}
~{\rm point}$ $ \rightarrow
 {\rm Infrared} ~ {\rm fixed} ~ {\rm point}$.

This formalism is similar to the 
Closed-Time-Path (CTP) formalism used in non-equilibrium 
quantum field theories~\cite{ctp}. The absence of
factorization is linked to the evolution from
a pure state $\vert A, B \rangle$ to a mixed density matrix, $\rho$,  
which cannot be described by a conventional $S$ matrix.

In our approach, the logarithmic algebra of the recoil 
operators forces 
the regularizing parameter $\epsilon$ (\ref{defeps})
to be identified with the logarithm of the
world-sheet area scale 
$A=\left|L/a \right|^2$, and hence with the target time. 
In the bounce picture outlined above, there will be a `breathing mode'
in the recoil-induced space-time, {\it characterized by
two directions of time}, 
corresponding to the processes of expansion, stasis and 
shrinking of the horizon in the recoil-induced space-time
(\ref{fourdime2}), all
within a few Planckian times. 
This is the Liouville-string
description of Hawking radiation.

\section{Physical Consequences of the Model}

\subsection{Modified Dispersion relations and Analogies with superfluids}

The result (\ref{refind}) implies a {\it modified dispersion relation} 
for matter propagation in the above-model of space time foam, which 
violates {\it linearly} Lorentz invariance on the brane D3
(we remind the reader that violation of Lorentz symmetry (LIV) 
is a generic feature of the recoil formalism we discuss here, 
and already occurs
in the bulk, as discussed in section 2.2):
\begin{equation} 
E^2-p^2=m^2 + p^2f(\frac{p}{M_{QG}})
\label{moddisp}
\end{equation}
where $M_{QG}$ denotes an effective scale at which quantum-gravitational
interactions set in. 
We note that such modifications of dispersion relations appear
as a generic feature of the non-critical string theory approach to 
quantum gravity, where the time is identified as the Liouville 
mode~\cite{aemn}. 

In the specific case discussed here (\ref{refrindex}) 
the function $f(\frac{p}{M_{QG}})$ is linear in the small quantity 
$p/M_{QG}$, where $M_{QG} \sim \frac{36}{285 g_s^2}M_D $.    
In general, other models of quantum gravity foam 
may yield more suppressed modifications,
where the function $f(p/M_{QG})$ starts from quadratic (or higher-order) 
terms in $p/M_{QG}$~\cite{mestres}. 
Clearly such modified dispersion relations are direct consequences
of the induced violations of Lorentz symmetry. Some experimental 
tests will be discussed later on. 

For the moment we note that~\cite{volkov}  
such modifications of dispersion relation is a typical effect occuring 
in open systems in condensed matter. In fact the non-diagonal 
form of the original induced metric (\ref{gemetric}) 
is common in condensed-matter situations where one 
encounters the motion of fermions in superfluids
in the Landau-Khalatnikov two-fluid framework~\cite{volovik}. 
We now turn briefly to a discussion of this analogy. 

Indeed, it was observed in~\cite{volovik} that 
relativistic fermionic quasiparticle 
excitations appear near the nodes of such systems, with a spin-triplet
pairing potential $V_{{\vec p}, {\vec p}'} \propto {\vec p} \cdot {\vec p}'$
and an energy gap function 
$\Delta ({\vec p}) \sim c p_x$ in the polar phase,
where $p_x$ denotes the momentum component along, say, the $x$ direction,
and $c$ denotes the effective `speed of light' in the problem. 
This is, in general, a function of the superflow velocity $w$: $c(w)$,
that is
determined self-consistently by solving the Schwinger-Dyson-type
equations that minimize the effective action.

This system was considered in the context of ${}^3$He 
in a container with stationary rigid walls and
a superflow velocity  $w$ taken, for simplicity, also along the $x$
direction.
The Doppler-shifted energy of the fermions in the pair-correlated
state with potential $V_{{\vec p}, {\vec p}'}$ is given by 
\begin{equation}
       E(p_x, \epsilon _{p} ) = \sqrt{\epsilon _{p} ^2 + c^2 p_x^2 }
+ wp_x ,
\label{he3}
\end{equation} 
where $\epsilon _p = (p^2 - p_F^2)/2m$, is the energy 
of the fermion in the absence of the pair correlation,
$p_F$ is the Fermi momentum
and $m$ is the mass of a superfluid (e.g. Helium) atom~\cite{volovik}. 
The term $w p_x$
appearing in
the quasiparticle energy spectrum (\ref{he3}),
as a result of the motion of the superfluid,  
yields an effective off-diagonal 
(1+1)-dimensional metric $G_{\mu\nu}$  
with (contravariant) components
\begin{equation} 
   G^{00}=-1, \qquad G^{01} = w , \qquad G^{11}=c^2 - w^2
\label{effhe3metric}
\end{equation} 
The off-diagonal elements of the 
induced metric 
(\ref{effhe3metric})
are analogous to those of our metric 
(\ref{gemetric}) upon the 
interpretation of the Liouville field as target time~\cite{emn98,volkov}. 
In this analogy,
the role of the recoil velocity $\vec u$ in our 
quantum-gravitational case is played by
the superflow velocity field $w$. 

An important feature of the superfluid case 
is
the appearance of an horizon 
that characterizes the metric (\ref{effhe3metric}).
This arises when the
superflow velocity $w = c$, in which case the metric element $G^{11}$ 
in (\ref{effhe3metric})
crosses zero, leading to a signature change for superluminal 
flow $w > c$. In fact, as shown in~\cite{volovik} by 
an analysis of the gap equation, the superluminal flow branch 
is not stable, because it corresponds to a saddle point rather than a
minimum
of the effective action. This suggests that the intactness of the analogy
with our
problem, in which on the one hand we do have the formation of horizons
(\ref{encon}), and 
on the other the special dynamics that governs the recoil 
problem~\cite{mavro+szabo} keeps the photon velocity
subluminal (\ref{refrindex}), may be maintained.

\subsection{Other Stochastic Effects: light-cone fluctuations} 

The above-mentioned modified dispersion relation (\ref{moddisp}) 
and the induced 
refractive index (\ref{refind}) may be viewed as a sort of {\it mean field 
effects} in the full quantum theory.
As discussed in \cite{ellis99}, the summation over world-sheet
topologies of the stringy $\sigma$-model we are examining here, 
leads to additional stochastic fluctuations of the widths
of the pulses of massless particle probes (photons) propagating in the 
space-time-foam background. Such fluctuations are associated with 
non-zero elements of the following (target-space) correlation functions
between two gravitons:
\begin{equation}
          \langle h_{\mu\nu} h_{\alpha\beta} \rangle _{QG} \ne 0 
\label{stochfluct}
\end{equation} 
where the correlators are taken with respect to the full theory 
of quantum-gravity foam. In our $\sigma$-model D-brane 
framework, 
to leading order such fluctuations 
incorporate resummation of a sub-class of           
world-sheet topologies (annuli)~\cite{mavro+szabo}, 
but in more complete situations other topologies must
be included, which makes the full expression unknown.  
Nevertheless, from the leading order calculations 
we have the result that such stochastic fluctuations 
will lead to a stochastic spread on the arrival times
of photons (massless probes) with the {\it same energy}, 
in contrast to the the (mean-field) refractive 
effect (\ref{refind}) which relates photons
in different energy channels. Moreover,  
the stochastic effect will be suppressed
by extra powers of the string coupling $g_s$ as 
compared to (\ref{finaldelay}).  
This phenomenon is similar 
with the one predicted in the context of conventional quantum-field
theory of gravity involving graviton coherent states~\cite{ford}
and is probed independently of any possible 
modification of dispersion relations. 

\subsection{Charged Particles and Transition Radiation?} 

As we have stressed above, 
the basic feature of our model~\cite{horizons} is the formation of `bubbles' 
(see fig. \ref{figbubble}) with non-trivial refractive index 
(\ref{refind}), thereby giving the notion of a `medium'. 
So far we have examined the propagation of massless neutral probes in 
such a medium. When charged particles are considered  
one is prompted to draw an analogy~\cite{efmmn} with the situation encountered
in electrodynamics of interfaces between two media with different
dielectric constants (and refractive indices). 
In such a situation the 
phenomenon of {\it transition radiation} (TR) takes place~\cite{jackson}:
when a charged particle crosses the interface separating
two media with different refractive indices and dielectric constants
radiation is emitted in the forward direction, which is appreciable 
for highly energetic particles. The physical reason for TR 
is the fact that the moving electromagnetic fields of the charged particle
induce a time-dependent polarization in the medium which emits radiation. 
The radiated fields from different points in space combine {\it coherently} 
in the neighbourhood of the path and for a certain depth (formation depth) 
in the medium, giving rise to TR with a characteristic 
angular distribution and intensity that depends on the Lorentz factor
(and hence the energy) of the charged particle. For 
the case of relevance to us here, in which the particle 
crosses an interface 
separating the vacuum from a medium with refractive index close to unity,
the bulk of the TR spectrum comprises of highly energetic photons. 

In the case of the space-time foam model discussed in \cite{horizons}
and reviewed here, one should expect similar effects if he/she 
views the D-brane defects as ``real''
(some sort of ``material reference system''~\cite{emncosmol,rovelli}), 
the recoil of which gives 
rise to the `physical' bubble picture of fig. \ref{figbubble}. 
Of course this is only an analogy,
and the actual calculation differs  from 
the electromagnetic case, 
especially due to the microscopic (Planckian) size of the bubbles,
which approaches the uncertainty limits.
However, we believe that qualitatively similar 
phenomena take place, but we expect them to be suppressed,
due to trapping of a significant part of the emitted radiation
inside the microscopic fluctuating horizons. 
It goes without saying that the characteristics
of the associated TR, if any, depend crucially on the {\it details}
of the space-time foam picture. It may even be averaged out.
Nevertheless, the possibility of detection of energetic photons
accompanying charged particles due to 
space time foam quantum-gravity effects is an interesting issue
deserving a separate study, which we shall turn to in a forthcoming
publication.

\subsection{Time-dependent ``vacuum'' energy} 

An important feature of our (non-critical) stringy model 
is the appearance of a non-trivial ``vacuum'' energy,
which actually is time dependent, relaxing to zero asymptotically. 
This is a generic feature of D-brane space-time models~\cite{emncosmol}
and may have important physical consequences. 
The recoil of the D-particles in our picture, as a result of 
the scattering of strings on them, {\it excites} the ground state 
of the system, with the inevitable consequence of the appearance 
of a non-zero (time-dependent) excitation energy, which plays the r\^ole of a
 time-dependent  ``vacuum energy''. Although the characterization
``vacuum'' is really misleading in the sense that in the recoil picture
one is dealing with an excited state of the D-brane system, 
however for our purposes
we shall continue to use it. This is on account of 
two facts: (i)   
in the framework of our model, 
an observer living on the recoiling D3 brane of fig. \ref{fig3} 
cannot tell the difference between living in an excited state 
and in the ground state of the system. (ii)  
in our scenario for the space-time foam the recoil
may represent {\it virtual quantum fluctuations}, which can be 
attributed to properties of the ``space-time-foam'' non-equilibrium 
state.

It is important to notice that in the situation depicted in 
fig. \ref{fig3} one encounters two kinds of contributions
to the ``vacuum'' energy. The first occurs as a result 
of recoil effects from the D-particles embedded in the D3 brane.
As discussed in \cite{emncosmol}, such effects lead to {\it positive} 
excitation (or ``vacuum'') energies exhibiting a $1/t^2$
time dependence as $t \rightarrow \infty$.  
In addition to this, one encounters contributions to the ``vacuum''
energy coming from the bulk space time, as a result of the recoil 
(quantum) fluctuations of the 
D3 brane along the bulk directions. Such contributions 
may lead to {\it negative} (anti-de-Sitter type)
contributions to the vacuum energy on the brane~\cite{adrian+mavro},
which in certain models also scale as $1/t^2$.

{\it Both} types of contributions to the vacuum energy 
are responsible for a supersymmetry {\it obstruction},
in the sense that the excited state 
of the system of 
recoiling D-branes (non-critical strings)  
is not supersymmetric, despite the fact that
the true ground state of the system (no recoil, critical strings) is. 
The possibility of opposite sign contributions may imply cancellations
in certain model yielding strongly suppressed or even zero vacuum energy. 
However, it should be remarked that a small 
positive contribution to the vacuum energy, scaling as $1/t^2$, 
is still compatible with (if not desirable on account of) recent 
astrophysical data, coming, for instance, from high-redshift
supernovae~\cite{sndata}.

\section{Experimental Tests} 

\subsection{Astrophysical Tests of Modified Dispersion Relations and 
the associated Lorentz Symmetry Violation} 

It is interesting to remark that 
the modified dispersion relations (\ref{moddisp}) 
may have falsifiable consequences in the foreseeable future,
especially if one looks at distant astrophysical probes
such as light from Gamma-Ray-Bursters (GRB)~\cite{sarkar,bilkar} 
and ultra-high-energy cosmic rays~\cite{olinto}. 
Such tests fall into the general category of 
experimental probes of Lorentz symmetry~\cite{lorentz,mestres,pullin}.
Indeed, in our framework Lorentz Invariance is only a symmetry of the 
{\it low energy} effective theory, which is reminiscent of the 
situation encountered in condensed-matter systems with relativistic excitations
near certain points (nodes) of their fermi surfaces (e.g. 
$d$-wave superconductors).
Near these points, the effective low-energy field theory appears 
relativistic, which is not the case of the full theory. 

In a similar manner, for energies much lower than the Planck energy, $M_P$,
which is the characteristic scale of Quantum Gravity, 
Lorentz symmetry seems a good symmetry of the effective theory,
whilst there may be violations of it for energies near $M_P$
where the concept of space and time may change drastically.
In the model considered above, we have seen that this is indeed the case,
and such manifestations of Lorentz Invariance Violations (LIV)  
are the modified dispersion relations (\ref{moddisp}) and the 
induced refractive index effects (\ref{refind}). 
We should notice, however, that the LIV in our model are different
from the ones suggested by Coleman and Glashow~\cite{lorentz}.
In that model the violations of Lorentz symmetry result in
the propagation 
velocity of massless particle species (e.g. neutrinos ) being species 
dependent $c_i$, whilst 
in our case (\ref{refind}) 
the violation depends only on the energy content
of the particle, being otherwise universal, 
and in this sense the 
induced LIV are compatible with the equivalence principle.

\begin{figure}[b] 
\begin{center}
\includegraphics[width=.4\textwidth]{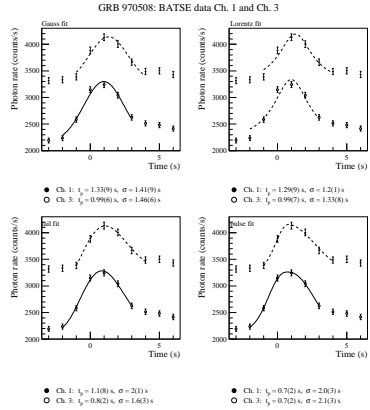}
\end{center}
\caption[]{Time distribution 
of the number of photons 
observed by BATSE in Channels 1 and 3 for GRB~970508, compared 
with the following fitting functions~\cite{efmmn}: (a) Gaussian, (b)
Lorentzian, (c) `tail' function, and (d) `pulse' function.
We list below each panel the positions $t_p$ and widths $\sigma_p$
(with statistical errors) found for each peak in each fit. 
The various fitting functions provide systematic errors in precision 
tests of quantum-gravity effects~\cite{efmmn}.}
\label{figdat1}
\end{figure}

\subsubsection{Gamma-Ray-Burster Observations} 

We presented in~\cite{efmmn} a detailed 
analysis of the astrophysical data for 
a sample of 
Gamma Ray Bursters (GRB) 
whose redshifts $z$ are known (see fig. \ref{figdat1}
for the data of a typical burst: GRB~970508).
We looked (without success) 
for a correlation with the redshift,
calculating a regression measure for the 
effect (\ref{finaldelay}) and
its stochastic counterpart associated with (\ref{stochfluct}). 
Specifically, we looked for linear dependences of
the `observed' $\Delta t$ and the spread 
$\Delta \sigma/E$ (with $E$ denoting energies) on ${\tilde z}
\equiv 2 \cdot [ 1 - (1 / (1 + z)^{1/2}] \simeq z - (3/4) z^2 +
\dots$, which expresses the cosmic-expansion-corrected redshift. 
We determined limits on
the respective quantum gravity scales $M_{QG}$ and $M_{{\rm stoch}}$ 
by constraining the possible magnitudes of the slopes  in 
linear-regression analyses of the differences between the arrival times
and widths of pulses in different energy ranges from five GRBs with
measured redshifts,
as functions of ${\tilde z}$.
Using the current value for the Hubble expansion parameter,
$H_0 = 100 \cdot h_0$~km/s/Mpc, where $0.6 < h_0 < 0.8$,
we obtained 
the following limits~\cite{efmmn}
\begin{equation} 
M_{QG} \ge 10 ^{15} \; {\rm GeV}, \quad M_{{\rm stoch}} \ge 2 \times 10 ^{15} \;
{\rm GeV} 
\label{limit}
\end{equation} 
on the possible quantum-gravity effects.

This is one kind of tests that could yield useful limits on 
space-time foam models in the future. However, I should remark
that such analyses have to be performed with care.
The regression index should yield reliable information 
only in case one has a statistically-significant population
of data, with known redshifts, something which at present 
is not feasible (it is worthy of pointing out, however, 
recent claims~\cite{fenimore} according to which a systematic study of the 
available luminosity and spectral lag 
data of GRB's can lead to an estimate of the respective redshifts.
If true, such studies can provide us with statistically significant samples
of GRB data relevant to our quantum-gravity precision tests).  
Moreover, detailed knowledge on the 
emission mechanisms~\cite{pirangrb} at the source are essential 
in order to disentangle effects that may be due to 
conventional physics, i.e. physics which is unrelated to 
foam effects. For instance, it is known~\cite{speed}
that non-trivial vacua in effective (non-linear) 
theories of quantum electrodynamics, associated with 
thermalized fermions or photons (e.g. cosmic microwave 
background radiation), lead to non-trivial refractive 
indices. However, the energy dependence on the probe energy 
in such cases 
is different from our effect (\ref{refind}), in the sense 
that it either leads to an energy-independent light velocity,
which simply changes value (jumps) as the Universe expands,
or it leads to an energy and temperature dependent refractive 
index, which however decreases with increasing 
probe energy, and hence leads to the opposite effect
than (\ref{finaldelay}). 
Such conventional physics effects, therefore, 
have to be disentangled from the 
pure space-time foam effects in all the relevant analyses.
It is probably worthy of mentioning that a systematic study of 
the observed GRB
indicates~\cite{norris} that the pulses of 
light become narrower, and the arrival times 
shorter, as one
goes from the low- to the high- energy channels. This is 
opposite to the quantum-gravity effect (\ref{finaldelay}),
and is probably related to conventional physics phenomena at the source. 

\subsubsection{Ultra-High-Energy Cosmic Rays} 

Recently, ultra-high-energy cosmic rays (UHECR) ~\cite{olinto},
with energies higher than $10^{20}$ eV,  
have been invoked~\cite{kifune,grillo,meyer,piran} 
as very sensitive probes
of Lorentz invariance violations, and in particular 
of the modified dispersion relations (\ref{moddisp}),
with sensitivities that, depending on the model, 
reach and/or exceed by far Planckian energy scales.
In particular, it has been argued~\cite{grillo} that
possible modifications of the dispersion relation,
due to quantum gravity effects,
could provide an explanation of the 
observed violation of the GZK cut-off~\cite{gzk}, and 
also be responsible 
for a significant increase in the 
transparency of the Universe~\cite{kifune}, 
in such a way so as
the sources of UHECR could be extragalactic,
lying much further (at cosmological
distances), in contrast with  
the common belief, based on Lorentz-invariant models, 
that the origin of UHECR 
should be within at most 50 Mpc radius from us~\cite{olinto}. 
Such modifications in the dispersion relation have also been 
invoked as a possible explanation of certain discrepancies 
between the 
observed $\gamma$ spectrum of Markarian 501 and expectations
based on a new estimate of the infrared background~\cite{meyer}. 

Although, most likely, conventional explanations could account for 
such violations of GZK cutoff and other UHECR 
related effects~\cite{fargion},
however the fact that 
such phenomena exhibit sensitivities to Planckian (and even 
sub-Planckian) scales is by itself remarkable and one cannot 
exclude the possibility of even having 
experimental signatures of quantum-gravitational 
effects in the near future.
For instance, it has been argued~\cite{piran} that certain models
of deformed Lorentz symmetry cannot lead to threshold effects that account 
for a violation of GZK cut-off, in contrast to our model, 
and hence UHECR data can 
be used to disentangle various models of space-time foam. 
In addition, since UHECR involve charged particles, the possibility 
of foam-induced 
transition radiation discussed in the previous section,  
should be taken into account as a means of excluding models.

\subsection{Terrestrial Experiments on space-time foam} 

We shall conclude this section by mentioning briefly that 
space-time quantum-gravity models can also be tested in 
terrestrial particle-physics experiments, 
involving neutral kaons~\cite{ellis84,ELMN,huet} and other sensitive probes
of quantum mechanics, as well as gravity-wave interferometric 
experiments~\cite{amelino}. 

In the latter case, the effects are similar to the ones associated
with the GRB experiment, and stem from the fact that space-time
foam
effects induce stochastic modifications 
of the dispersion relations for photons, and as 
such appear as noise in the gravity-wave 
interferometers. It is remarkable that the next-generation
facilities of this kind could be sensitive (in principle) to 
Planck-size effects~\cite{amelino}.

In the neutral-kaon or other meson 
experiments, the 
effects of quantum gravity 
are more subtle and are associated with 
the modification of the quantum mechanical evolution 
of the density matrix of (neutral kaon) matter
propagating in a space-time foam background.  
A parametrization of possible deviations
from the Schr\"odinger evolution  has been given~\cite{ellis84},
assuming energy and probability conservation, in terms of
quantities $\alpha,\beta,\gamma$ that
must obey the conditions
$\alpha , \, \gamma \, > \, 0, \qquad \alpha \gamma \, > \, \beta^2$
stemming from the positivity of the density matrix $\rho$. 
These parameters 
induce quantum decoherence and violate CPT~\cite{ELMN}. 
Experimental data on neutral kaon decays so far agree perfectly with
conventional
quantum mechanics, imposing only the following upper limits~\cite{emncplear}:
\begin{equation}
\alpha < 4.0 \times 10^{-17} \hbox{GeV}, \qquad 
\beta < 2.3 \times 10^{-19} \hbox{GeV}, \qquad 
\gamma < 3.7 \times 10^{-21} \hbox{GeV}
\label{bounds}
\end{equation}
We cannot help being impressed that these bounds are in the
ballpark of $m_K^2 / M_P$, which is the maximum 
magnitude that we could expect any such effect to have.

It has been 
pointed out in ref. \cite{benati} that within the context of linear maps
the requirement of complete positivity of $\rho$ results in 
further restrictions, in particular only two of the coefficients,
$\alpha,\gamma$  
are non vanishing. Together with energy conservation,
this requirement leads to the so-called Lindblad parametrization
for the non-quantum mechanical evolution~\cite{lindblad}. 
Recently, however, it has been argued~\cite{adler} that     
such a special form of linear evolution leads to effects
that are significantly more suppressed than (\ref{bounds}),
by many orders of magnitude, which casts doubt on 
claims~\cite{benati} that complete positivity of linear maps
can be tested experimentally in the foreseeable future. 

However, in our Liouville-string case, which is the basis of
the present model of quantum gravity foam, 
the deviations from the conventional quantum mechanical evolution 
in the respective density matrix are {\it non linear}~\cite{emn98,emnadler}.
In such a case, the parameters $\alpha,\beta,\gamma$ can only be 
viewed as appropriate averages, which themselves depend on the wave-functional
of the system. 
This is a non-trivial feature of our formalism of identifying 
time with the Liouville mode~\cite{emn98}. In the particular case 
of recoiling D-particles the non-linear temporal evolution 
of the system has been  
computed explicitly~\cite{mavro+szabo2}. Such non-linear evolution 
of the quantum-gravitational `environment' also results, in general, 
in a 
non-linear evolution of low-energy matter propagating 
in this background.
Moreover, in this specific example energy is not conserved
during the scattering of strings with D-particle defects, 
as a result of recoil of the latter. 
This is also in agreement 
with the explicit breaking 
of translational invariance by the presence of D-branes,
and is captured formally by the 
very special properties of Liouville dynamics. 

As discussed in \cite{emnadler}, such features lead
to important deviations 
from the (specific) Lindblad evolutionary form.
which, in turn implies the possibility of significant enhancement 
of the effects: the latter can be as large as (\ref{bounds}), thereby 
offering the possibility of experimental 
tests in the next-generation of neutral-kaon (or other meson) 
experiments~\cite{ELMN,huet}, 
as well as future neutrino facilities~\cite{lisi}.

\section{Instead of Conclusions} 

We have discussed in this article a microscopic mechanism for the 
dynamical formation of
horizons by the collisions of closed-string
particle `probes' with recoiling D--particle
defects embedded in a $p$-dimensional 
space time, which may in turn be viewed as
a $Dp$ brane domain wall in a higher-dimensional target space. 
As we have argued before,
the correct incorporation of recoil effects, which are unavoidable 
in any quantum theory of gravity that reproduces the 
conceptual framework of general relativity in the classical limit, 
necessitates a Liouville string approach in the context
of a (perturbative) world-sheet framework. 

The most important result of our approach is the
demonstration of the dynamical formation of breathing horizons, 
which follows directly from the 
restoration of conformal invariance by means
of Liouville dressing. 
The non-trivial optical properties induced by
the propagation of light in such a fluctuating space-time
may be subject to experimental verification 
in the foreseeable future, and are already
constrained by existing data~\cite{sarkar,efmmn}. 
The fact that the refractive index in the bubbles of
space-time foam is subluminal implies the absence 
of birefringence in light propagation, which is, however,
possible in other approaches to space-time foam~\cite{pullin}. 

It is important to stress that the sensitivity to Planckian 
effects may not be so remote as one naively thinks.
There are both terrestrial and astrophysical experimental tests,
which are currently under way or about to be launched, 
that may not be far from excluding (or even verifying!)
space-time foamy models of quantum gravity.
It goes without saying that such a sensitivity is 
highly model dependent. For instance the sensitivity of the GRB 
test to the effect (\ref{finaldelay}) 
depends crucially on the specific model 
of foam described above, in which the effect is 
linearly suppressed by the quantum gravity scale.
If, for some reason, the effect is quadratically 
suppressed~\cite{mestres}, 
such a sensitivity is lost~\cite{efmmn},
however such quadratic models may be experimentally testable by 
means of UHECR experiments. Within the framework of our model,
the linear effect is undoubtedly a feature of the single scattering 
event of a string with a D-particle in the foam.
In the case of many D-particles 
it is possible that the effect is further suppressed. 
This depend on the (yet unknown) details of the 
statistical dynamics of the foam.

Certainly much more work, both theoretical and `phenomenological'
is necessary before even tentative conclusions are reached on 
such important matters as an understanding and the possibility of
experimental signatures of quantum gravity. 
But as explained above, there are optimistic signs that this task 
may not be impossible. Allow me, therefore, 
to close this lecture by recalling a statement 
from Sherlock Holmes (by Sir A. Conan-Doyle), which was reminded to me 
by my collaborator A. Campbell-Smith: 
{\it If you eliminate the impossible, whatever remains, no matter
how improbable, must be the truth}. 

\section*{Acknowledgements}

This work is dedicated to the memory of T. Ypsilantis.
The work of N.E.M. is supported in part by PPARC (UK).
The author wishes to thank Prof. H. Klapdor-Kleingrothaus (Chairman) 
and
Dr. B. Majorovits
for inviting him to the DARK2000 third international conference
on Dark Matter in astro- and particle physics 
(Heidelberg, Germany, July 10-15 2000).
He also thanks  
D. Fargion and V. Mitsou for discussions on cosmic rays and  
transition radiation respectively, and  
H. Hofer for his interest and support.

\end{document}